# How Polarized Have We Become? A Multimodal Classification of Trump Followers and Clinton Followers

Yu Wang, Yang Feng, Zhe Hong, Ryan Berger and Jiebo Luo

University of Rochester
Rochester, NY, 14627, USA

**Abstract.** Polarization in American politics has been extensively documented and analyzed for decades, and the phenomenon became all the more apparent during the 2016 presidential election, where Trump and Clinton depicted two radically different pictures of America. Inspired by this gaping polarization and the extensive utilization of Twitter during the 2016 presidential campaign, in this paper we take the first step in measuring polarization in social media and we attempt to predict individuals' Twitter following behavior through analyzing ones' everyday tweets, profile images and posted pictures. As such, we treat polarization as a classification problem and study to what extent Trump followers and Clinton followers on Twitter can be distinguished, which in turn serves as a metric of polarization in general. We apply LSTM to processing tweet features and we extract visual features using the VGG neural network. Integrating these two sets of features boosts the overall performance. We are able to achieve an accuracy of 69%, suggesting that the high degree of polarization recorded in the literature has started to manifest itself in social media as well.

## 1 Introduction

> To see a World in a Grain of Sand
> And a Heaven in a Wild Flower,
> Hold Infinity in the palm of your hand.
> And Eternity in an hour.
> –William Blake

Polarization in American politics has been extensively documented and analyzed for decades, and it became all the more apparent in the 2016 presidential election, with Trump and Clinton depicting two radically different pictures of America and targeting "two different Americas".[1] According to a survey by the Pew Research Center, only 8% of Democrats and Democratic-leaning independents approve of Trump's job performance during the first month of his presidency, as compared to 84% approval rate by Republicans and Republican leaners.[2] A better understanding of the polarization phenomenon is urgently called for.

---

[1] http://www.denverpost.com/2016/11/04/clinton-derides-trumps-fitness-he-disparages-her-honesty.

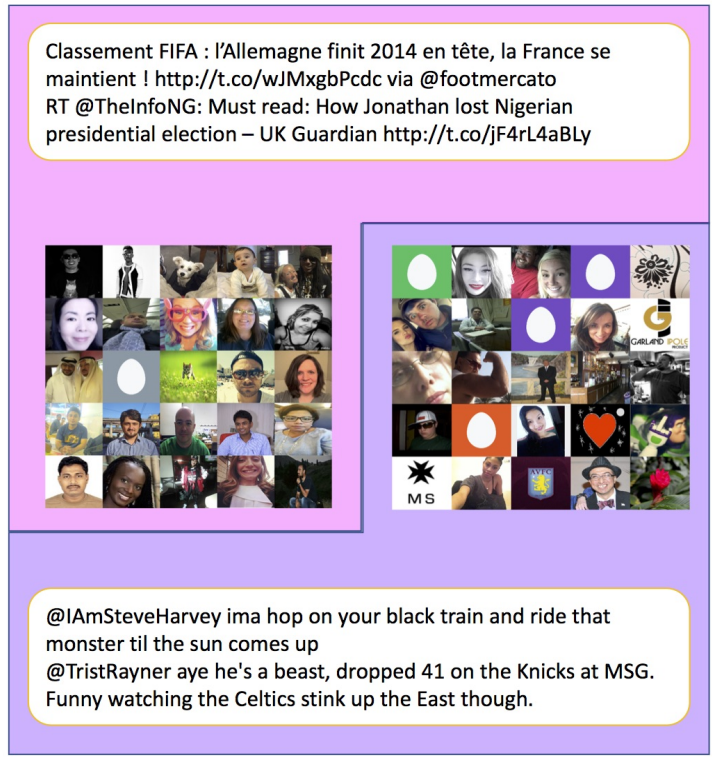

Fig. 1: Top left: profile images and tweets from Clinton followers. Bottom right: profile images and tweets from Trump followers. These everyday details constitute the source from which we detect polarization.

Twitter played an important role in the 2016 U.S. presidential election. All the major candidates including Donald Trump, Hillary Clinton, Bernie Sanders and Ted Cruz, have millions of Twitter followers, to whom they can easily reach out to. Candidates formulate issue policies, attack rival candidates and gauge the public opinions via 'likes' [34] over Twitter. Some of the candidates' tweets have even entered into the Democratic and the Republican debates. After winning the election, Donald Trump himself commented that tweeting "is a great way of communication" [30]. In his book *Our Revolution*, which reflects on the 2016 presidential campaign, Bernie Sanders suggests that one of the reasons why his campaign did well is the campaign team's success with social media [28]. One opinion shared by both Trump and Sanders is that having a large number of followers on Twitter is an invaluable campaign asset [30, 28]. Naturally, being able to distinguish between Trump followers and Clinton followers and understand why individuals choose to follow one candidate but not the other becomes important.

---

[2] http://www.people-press.org/2017/02/16/in-first-month-views-of-trump-are-already-strongly-felt-deeply-polarized.

In this paper, we take a first step in analyzing polarization of the public on social media. We treat polarization as a classification problem and study to what extent Trump followers and Clinton followers on Twitter can be distinguished. Our work differs from previous research in two different ways. First, previous work relies on expert-designed questions that have explicit political intonation. Our work, by contrast, tries to gauge one's Twitter following inclination by sifting through their daily tweets, profile images and posted pictures. Second, previous studies focus exclusively on American citizens and rely on small-sized surveys. Our work, boosted by the international reach of social media, includes individuals from all around the world and is several magnitudes larger than any survey-based studies.

We apply LSTM to processing tweet features and we extract visual features using the VGG network. Combining two sets of features, our model attains an accuracy of 69%, indicating that the high degree of polarization recorded in surveys has already manifested itself in social media.

Our contributions can be summarized as follows:

- We analyze the everyday language and pictures of Trump followers and Clinton followers and examine their high-level differences.
- We study political polarization from a completely new, non-invasive and more challenging angle that at the same time covers a larger portion of the population.
- We have made our data publicly available to facilitate replication and stimulate further analysis.

## 2 Literature Review

Earlier works have studied the increasing polarization of American politics at both the elite level [10, 21] and the mass level [3, 5, 7]. Druckman *et al.*, in particular, study how elite partisan polarization affects public opinion formation and find that party polarization decreases the impact of substantive information [6]. Social clustering, the opposite side of polarization, is analyzed in [22, 1]. We contribute to this literature by analyzing polarization at the public level on the social media site Twitter.

There also exist a series of social media studies on the 2016 U.S. presidential election that are the inspiration of ours. The social demographics of Trump followers and Hillary followers are analyzed in [35] using exclusively profile images. Again using profile images, [33, 31] further tracked and compared the 'unfollow' behavior of Trump followers and Hillary followers. [37] and [40], on the other hand, analyze the tweets posted the presidential candidates and draw inference about the followers' preferences based on the observed number of 'likes.' Our work follows this line of research. Compared with previous studies, we utilize both text information and images.

Recursive neural networks (RNN), in particular its variant Long Short Memory Networks (LSTM) [12], have been very successful in modeling sequential data [43, 23], with wide applications in text classification [16], especially in sentiment analysis [14, 19]. A recent study [27] that is most related to ours uses word embeddings and LSTM to classify as either Democratic or Republican the social media messages posted by people who are known to be either Democrats or Republicans. Our study, by comparison, is more challenging as we attempt to capture political leaning among ordinary Trump/Clinton followers, who may not be Democrats or Republicans and may not even be American voters.

Deep convolutional neural networks (CNN), on the other hand, have proven extremely powerful in image classification, usually reaching or surpassing human performance [29, 11, 36]. There exist quite a few studies that analyze selfies and other images posted on the social network [4, 42]. In particular, researchers have attempted to capture demographic differences (e.g. gender and race) between Trump followers and Clinton followers by examining exclusively the profile images [35, 33]. Our work analyzes the same profile images to detect Twitter following inclinations from these images.

## 3  Problem Definition

We define the problem of measuring polarization on social media as a classification problem: whether the individual under study is following Trump or Clinton. The high-level formulation is

$$\Pr(\text{Predicted Class} = \text{Trump} \mid \text{Tweet}, \text{Picture}, \text{Trump follower})$$

$$\Pr(\text{Predicted Class} = \text{Clinton} \mid \text{Tweet}, \text{Picture}, \text{Clinton follower})$$

where **Tweet** and **Picture** are two sets of features and the classes of Trump and Clinton followers are balanced. Tweets and pictures that exhibit a low level of polarization will bring accuracy close to 0.5, whereas a high level of polarization on social media could push accuracy substantially over 0.5.

Note that *follower* is a rather weak concept: it implies neither support nor opposition. At a minimum, it suggests interest. We are acutely aware that polarization might be too strong a word to be associated with Twitter following which is technically just "one click away" [38]. Referring back to William Blake, "to see a world in a grain of sand," we strive to detect traces of polarization as millions of individuals cluster around two most polarized candidates.

## 4  Data and Preprocessing

On May 8th 2016, Donald Trump had 8.02 million followers and Hillary Clinton had 6.18 million followers. To collect ground-truth data, we first use binary search to identify and remove individuals who were following both candidates and second we randomly select 3905 individuals who are following Donald Trump but not following Hillary Clinton and 3668 individuals who are following Hillary Clinton but not Donald Trump. In order to make our classification task easier, we do not select individuals that were following both candidates[3]. We choose to collect data from a pre-election date because this will partially insulate our study from the effect of Trump winning the election and subsequently becoming the President. The data collection period is reported in Figure 2.

---

[3] Individuals who follow both candidates constitute a surprisingly small portion of the entire follower population. For a more detailed analysis that includes also Bernie Sanders, please see [38].

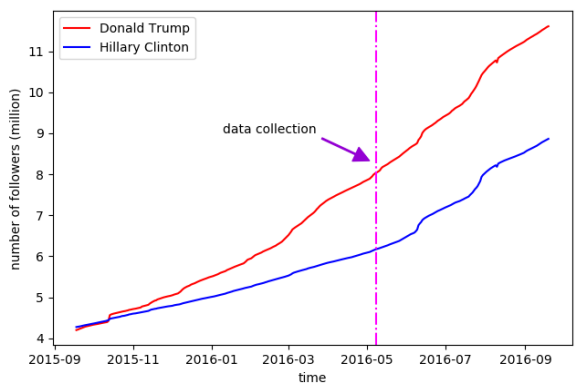

Fig. 2: We collected the IDs of the original seed individuals on May 8th, six months before the election day.

The selected individuals in turn serve as seed users, and we collect their profile images, tweets and posted pictures. While each individual could only have one profile image, the number of tweets and posted pictures is virtually limited. In Table 1, we present the summary statistics of the data used in this paper. Compared with questionnaire-based studies, which have been the norm, our approach is substantially more subtle and voluminous.

Table 1: Summary statistics

|                | Trump followers | Clinton followers |
|----------------|-----------------|-------------------|
| Individuals    | 3,905           | 3,668             |
| Tweets         | 2,214,898       | 2,172,591         |
| Profile Images | 3,096           | 2,956             |
| Posted Pictures| 110,000         | 110,000           |

### 4.1 Tweets

Tweets constitute an integral part of our data source. From these tweets we attempt to detect signs of Twitter following inclination. Compared with questionnaires where the surveyed subject takes a passive role and answers a limited number of carefully crafted questions, tweets represent the spontaneous flow of emotions, ideas, thoughts and events.

For example, in what follows we show two questions that are typically used for the study of political polarization [13].

---

Question 1:
A. Government is almost always wasteful and inefficient.
B. Government often does a better job than people give it credit for.
Question 2:
A. Poor people today have it easy because they can get government benefits without doing anything in return.
B. Poor people have hard lives because government benefits don't go far enough to help them live decently.

---

Answers to these questions are then transformed into binary values and their mean value is then used to represent an individual's political leaning between left and right.

By comparison, our text data are substantially subtler. Below are four tweets posted by followers of Donald Trump and Hillary Clinton:

---

@markhoppus how much you gettin an hour out there?
@wizkhalifa like when you finally release your load into the sock
RT @somizi: Topic at church: WISDOM. With it u will be able to survive anything anywhere
@allinwithchris My mom actually went to high school with Dick Cheney in Wyoming. So he is "from there" - just not Lynn. :-) (Weird)
@afc33125: OK ALREADY It's time to CUT OFF #Christie's FREE AIR TIME!!!!!!

---

To process these tweets, we first tokenize the text using Python's *nltk.tokenize* library which is able to preserve punctuations. An illustrative example is reported below:
Original tweet:
*Today is the anniversary of Congress' most deplorable act, an apt day ponder what a petty, partisan, tribal, prick @morning_joe is.* After tokenization:
*Today is the anniversary of Congress ' most deplorable act , an apt day ponder what a petty , partisan , tribal , prick @morning_joe is .*

The summary statistics of the tweets after tokenization is reported in Table 2 and the Figure 3 shows the distribution of these tweets for each group. It can be observed that while Donald Trump is usually thought of using short and easy-to-understand language, his tweets are statistically longer than Clinton's. Interestingly, the tweets posted by Clinton's followers tend to be longer than those of Trump's followers.

Second, we use a pre-trained word embedding [25, 15] to translate each token into an array of dimension 25 and 50 respectively. Word embeddings map words into a higher dimensional vectors that can capture syntactic and semantic patterns and have been widely used in text classification tasks [27, 18]. To overcome the limited length of a tweet, we concatenate every 20 tweets from the same individual into a super-tweet and in our final processing super-tweets that are shorter than 100 tokens are discarded. Eventually, we obtain 197,440 training examples ( the ratio is balanced with 97,528 for Clinton followers and 99,912 for Trump followers), and 10,000 examples for validation and 10,000 for testing.

Table 2: Distribution of Tweet Lengths

|  | Mean | Variance | # Observations |
|---|---|---|---|
| Clinton | 20.4 | 40.8 | 3234 |
| Clinton followers | 16.2 | 80.4 | 2,172,591 |
| Trump | 21.4 | 50.2 | 3602 |
| Trump followers | 14.7 | 69.6 | 2,214,898 |

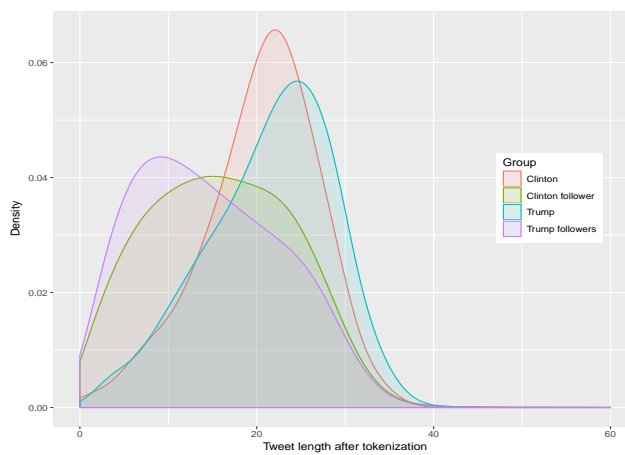

Fig. 3: The length distribution of tweets posted by Clinton, Clinton followers, Trump and Trump followers respectively.

### 4.2 Profile pictures

With the recent success of deep convolutional neural networks (CNN), profile pictures and selfies have emerged as an integral data source for understanding individuals' demographics, sentiments and habits [4, 42, 36]. In particular, several studies of the 2016 presidential election have demonstrated the efficacy of using images for modeling political events. [33] studies the unfollowing behavior of men and women, and it relies on a neural network to infer individuals' gender. So does [32], which studies the effects of the 'woman card' controversy.

As each one is confined to having only one profile image, we are able to collect 6,052 profile images. We will attempt to make predictions of one's Twitter following behavior based on the deep features in the these profile images.

### 4.3 Posted Pictures

Besides profile images, we also collect pictures posted by the seed individuals to train our integrated LSTM-CNN model. In total we have 220,000 posted pictures.

In order to have a better understanding of these images, we first use the VGG neural network to extract deep features. Then we use k-means to cluster these images and use the silhouette coefficient to set the proper number of clusters (i.e., k) to 27. We subsequently merge the clusters that we judge to be of the same category and eventually we get 24 clusters. For each cluster, we are able to label it according to the images therein. The names of these 24 clusters and the associated numbers of images are reported in Table 3.

Table 3: Cluster Labeling and Sizes

| ID | Cluster | # Images | ID | Cluster | # Images |
| --- | --- | --- | --- | --- | --- |
| 1 | people indoor | 3820 | 13 | screenshot | 1542 |
| 2 | text | 5462 | 14 | building | 1323 |
| 3 | leader & speaker | 1793 | 15 | dog & cat | 916 |
| 4 | sports | 1649 | 16 | round object | 1070 |
| 5 | soldier | 1604 | 17 | cartoon | 1820 |
| 6 | car & bus | 1086 | 18 | food | 699 |
| 7 | table | 1381 | 19 | scenery | 1233 |
| 8 | animal | 676 | 20 | poster | 1604 |
| 9 | selfie | 1475 | 21 | square object | 1473 |
| 10 | people outdoor | 1715 | 22 | suit & tie | 1463 |
| 11 | nature | 949 | 23 | other | 1738 |
| 12 | crowd | 1795 | 24 | female | 1714 |

We notice that the two dominant clusters are text and people indoor respectively and that they are several times larger than the smaller clusters such as food and animal. This indicates that Trump followers and Clinton followers are most likely to post pictures that fall into these two categories. Figure 4 shows that top clusters from our clustering result.

Besides the large variation in cluster sizes, we further explore with-in cluster differences between Trump followers and Clinton followers. In Figure 5, we show the distribution of images for Trump followers and Clinton followers separately. We observe, for example, more images fall into the sports cluster and the car cluster for Trump clusters. At the same time, more *crowd* and *cartoon* images are posted by Clinton followers than by Trump followers in relative terms.

To test for the statistical significance of these differences, we use score test [33] to each cluster. The test results suggest that Clinton followers have a statistically higher representation in *leader & speaker, table, crowd, catoon, suit & tie, and female*. Trump followers have a statistically higher representation in *sports, vehicle, screenshot, round object, food, and scenery*. Differences in the other clusters, including the top two clusters, *text* and *people indoor*, are not statistically significant. Figure 7 at the end of our paper summarizes our test results.

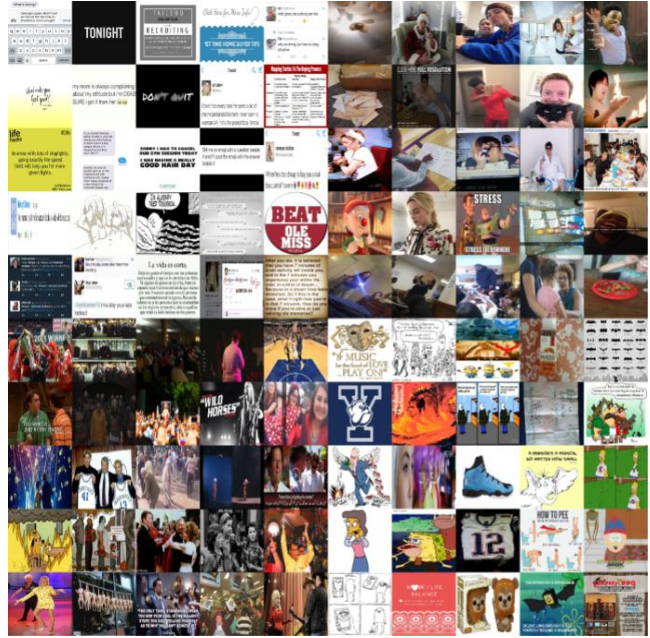

Fig. 4: Top left: text; top right: people indoor; bottom left: crowd; bottom right: cartoon.

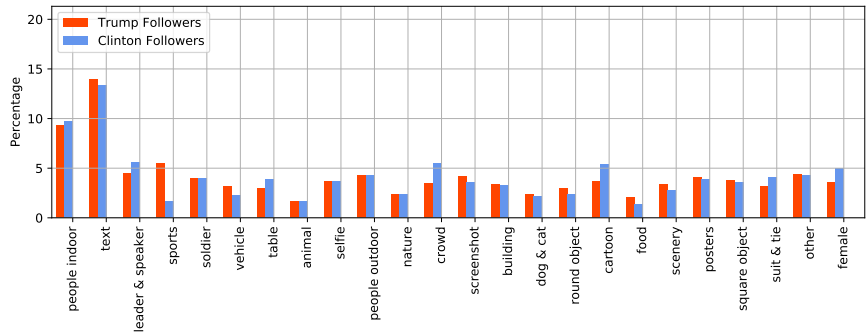

Fig. 5: The distribution of the 24 clusters for Trump followers and Clinton followers. Notice that the weight on the sports cluster is much higher for Trump followers than that for Clinton followers. The opposite holds for the crowd cluster, the cartoon cluster and the female cluster.

These variations in distribution among clusters offer us some first evidence that posted pictures can be useful for detecting Twitter following inclination.

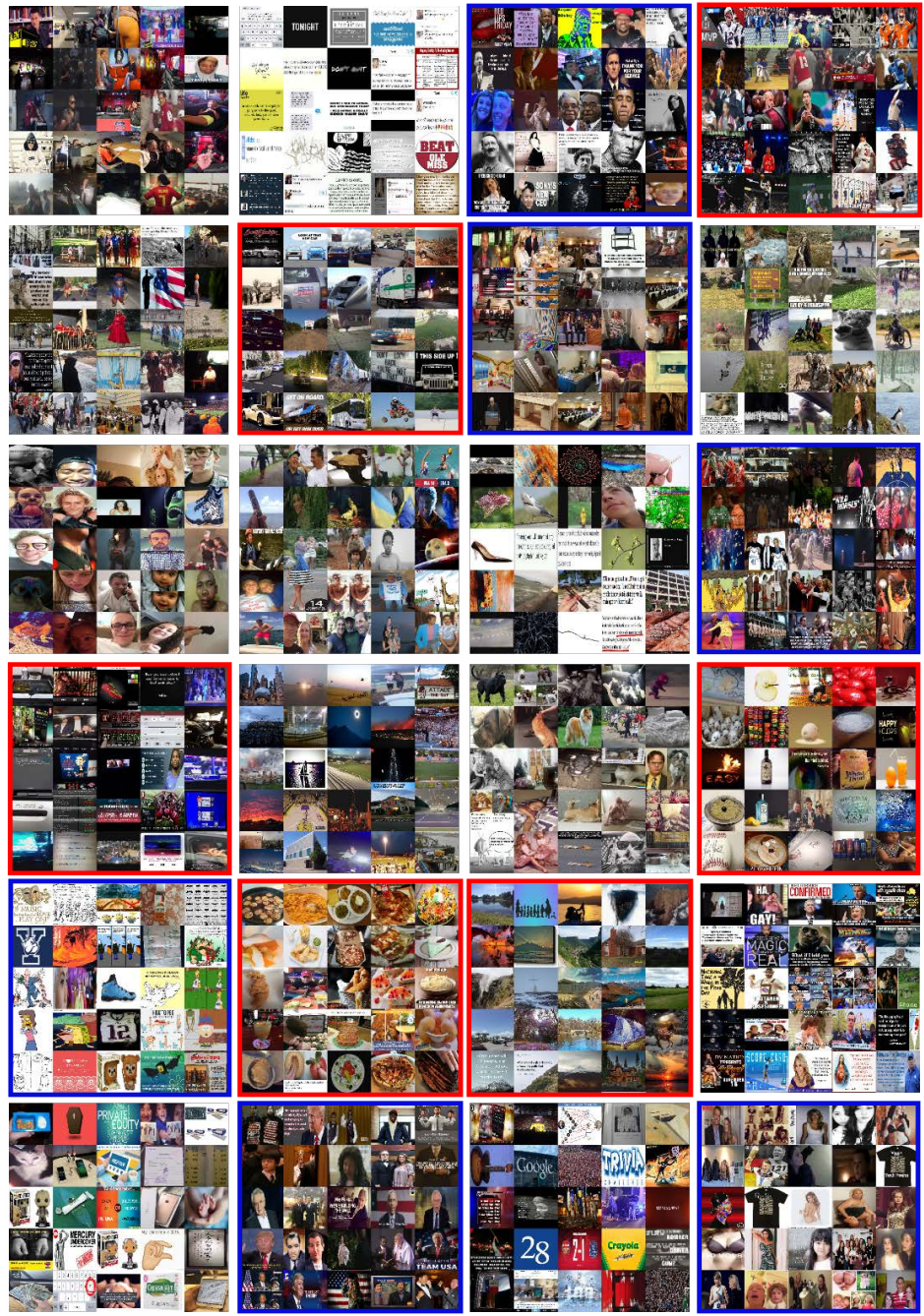

Fig. 6: Clusters within red bounding boxes have a higher representation among Trump followers. Clusters within blue bounding boxes are better represented among Clinton followers.

## 5 Experimental Setup

Our experimental design consists of four steps. First we classify tweets only, second we classify posted pictures, third we classify profile images, and fourth we integrate the two using a fusion model.

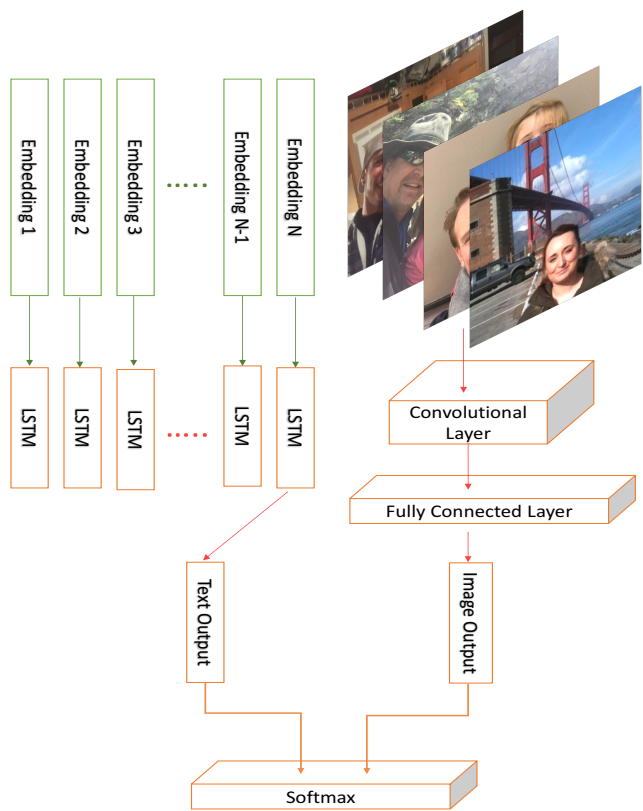

Fig. 7: The left part represents the LSTM component, which generates the text feature. The right part represents the CNN component, which generates the visual feature. The softmax function takes both features as input and generates the prediction.

### 5.1 Tweet - based Classification

We first use a pre-trained word embedding (with dimension $D$=25, 50) [25] to transform the tokens into word vectors. We then set the number of timesteps to 150. Super-tweets that were shorter than 150 are padded with vectors generated from a normal distribution with mean 0 and variance 0.1. For super-tweets longer than 150, we only take the first 150 word vectors. In this way we transform each super-tweet into a matrix of dimension D × 150.

These matrices in turn serve as the input of our LSTM network. To identify the best performing model, we set the dimension of the hidden state to 100 and 200 alternatively. To link the outputs of the LSTM cells to the class label, we experiment with (1) using the output of the last timestep, which has the same dimension as the hidden state, and (2) using the mean output of all the timesteps [9]. The highest accuracy achieved by this model is 67.9%.

Next we compare the performance of SVM with LSTM. Following [18, 41], we compute the vector representation of a super-tweet by averaging all the word vectors contained in the super-tweet. Given the padding and truncation operations discussed above, in effect, super-tweet (i) is represented as follows:

$$C_i = (\sum_{j=1}^{150} w_{i,j})/150$$

where $w_{i,j}$ is the word vector for the jth word in super-tweet i and 150 is the number of timesteps that we have previously chosen.

As with LSTM, we experiment with two different dimensions of the input: 25 and 50, both using the RBF kernel. For SVM we find that word representations of dimension 50 yield better performance. Nonetheless, LSTM outperforms SVM in all specifications. We report the details in Table 4.

### 5.2 Picture-based Classification

Given the large number of posted pictures in our dataset, we are able to finetune the entire VGG network rather than only the last few fully connected layers. We use 200,000 examples for training (class ratio 1:1), 10,000 samples for validation and another 10,000 samples for testing. The highest accuracy that we obtain is 59.4%.

In absolute terms, this accuracy is low and it reminds of the findings in picture clusters: there are a few clusters where Trump followers and Clinton followers weigh differently, but these are only minor clusters. For the dominant two clusters, *text* and *people indoor*, there is not statistical difference. In relative terms, however, this result is surprisingly high, as it confirms that it is feasible to predict individuals' Twitter following behavior by directly looking at their posted pictures.

Besides finetuning the entire architecture, we also experiment with extracting deep features using the VGG network and then feed these features into an RBF kernel SVM, reminiscent of the classical R-CNN architecture [8, 20]. This approach is particularly helpful when the number of training samples is small [24]. With this approach, we are able to achieve an accuracy of 57.0%, which is slightly lower than finetuning the whole architecture.

### 5.3 Profile Image - based Classification

Next, we examine the feasibility of predicting Twitter following inclinations using profile images. We have 3,096 profile images from Trump followers and 2,956 profile im-ages of Clinton followers. For each class, we take 500 samples out for validation and another 500 samples for testing.

As explained in the last subsection, we use the VGG network to extract deep features instead of finetuning it. Considering that the deep feature is of dimension 1,000, we further apply a maxpool operation of dimension 10 with stride 10 to the deep feature to obtain a final feature of length 100. We train an SVM using these features and the accuracy is 55.8% (Table 4).

### 5.4 Tweet & Picture - based Classification

Lastly, we develop an integrated approach that leverages both tweets and posted pictures. The model is illustrated in Figure 6. On the left is the LSTM model that processes the super-tweets, and on the right is the VGG model that processes pictures and generates deep features. At the bottom lies the softmax function that incorporates both text features and visual features:

$$\text{softmax}(W_{text}X_{text} + W_{visual}X_{visual} + Bias)$$

where $W_{text}$ and $W_{visual}$ represent two weight matrices that are to be learned from the data.

As with LSTM, we experiment with different configurations on the dimension of the word embedding, the dimension of the hidden state and the final output feature vector. The highest accuracy achieved by this fusion model is 68.9%, when the word vectors are of length 25, the hidden state is of dimension 100, and the mean output is used as the final output feature (Table 4). It should also be noted that across all specifications, the fusion model uniformly yields the best result.

Table 4: Experiment results

| Experiment Setup | Tweet | | Profile Image | Posted Picture | | Tweet & Picture |
|---|---|---|---|---|---|---|
| | SVM* | LSTM | SVM** | SVM** | VGG** | LSTM+VGG |
| Hidden 100, Embedding 25, Last | 61.1 | 67.8 | 55.8 | 57.0 | 59.4 | **68.5** |
| Hidden 100, Embedding 50, Last | 62.1 | 67.0 | 55.8 | 57.0 | 59.4 | **67.6** |
| Hidden 200, Embedding 25, Last | 61.1 | 67.5 | 55.8 | 57.0 | 59.4 | **68.6** |
| Hidden 200, Embedding 50, Last | 62.1 | 66.8 | 55.8 | 57.0 | 59.4 | **67.2** |
| Hidden 100, Embedding 25, Mean | 61.1 | 67.1 | 55.8 | 57.0 | 59.4 | **68.6** |
| Hidden 100, Embedding 50, Mean | 62.1 | 67.8 | 55.8 | 57.0 | 59.4 | **68.9** |
| Hidden 200, Embedding 25, Mean | 61.1 | 67.2 | 55.8 | 56.7 | 59.4 | **68.1** |
| Hidden 200, Embedding 50, Mean | 62.1 | 67.9 | 55.8 | 57.0 | 59.4 | **68.1** |

* For the SVM on tweets, we experiment with dimension of 25 and of 50. ** These models are not changed as we experiment with different LSTM specifications.

## 6   Discussion and future research

Individual attributes such as age, gender, sentiment and even income, can be estimated using either texts [2, 39, 26] or images [17, 42]. Both data sources certainly contain informative cues about these attributes. Our study suggests that these two approaches are at least capturing some shared hidden information, as the net gain of fusion model is less than than sum of the LSTM model and the VGG model.

Whiles texts and pictures both contribute to a better understanding of an individual's following inclinations, our experiments suggest that 4.4 million tweets contain more information than 220,000 pictures with regard to understanding Twitter following inclinations. Moreover, the best model is the one that uses both texts and pictures.

In terms of time complexity, the fusion model (LSTM+VGG) also outperforms fine-tuning VGG. For the fusion model, where deep visual features are extracted only once, training 20,000 samples for one epoch takes only about 9.3 minutes, compared with 2.8 hours for fine-tuning the VGG network for one epoch.

Polarization has been developing for decades, and now we have shown that the phenomenon has started to manifest itself in social media as well. A natural next step is measure and track the evolution of polarization in social media each year parallel with survey-based research. To facilitate this initiative, we have made our data and code available[4].

## 7   Conclusion

Inspired by the gaping polarization in our society and the increasing popularity of social media, we took a first step in measuring political polarization on Twitter. By assembling a large dataset and applying state-of-the-art algorithms, we achieved a high accuracy of 69% in predicting candidate followers. Methodologically, we demonstrated incorporating both text features and visual features could improve the model performance. Substantively, we provided first evidence that the gaping polarization observed in our society has crept into social media as well.

## 8   Acknowledge

We acknowledge support from the Department of Political Science at the University of Rochester, from the New York State through the Goergen Institute for Data Science, and from our corporate sponsors. We also thank the four anonymous reviewers for their insightful comments and suggestions.

---

[4] The datasets and codes are available at https://sites.google.com/site/wangyurochester/papers.